\documentclass[traditabstract]{aa}
\usepackage{txfonts}
\usepackage{graphicx}
\usepackage{subfigure}
\usepackage{color}
\usepackage{natbib}

\begin{document}

\title{The Halos and Environments of Nearby Galaxies (HERON) Survey IV:
Complexity in 
the boxy galaxies NGC 720 and NGC 2768} 

\author{
Andreas J. Koch-Hansen\inst{1} 
  \and Anna Pasquali\inst{1}
  \and R. Michael Rich\inst{2}
  \and Ortwin Gerhard\inst{3}
  \and Oliver M\"uller\inst{4}
  }
  
\authorrunning{A.J. Koch-Hansen et al.}
\titlerunning{Boxy galaxies NGC 720 and NGC 2768}
\institute{Zentrum f\"ur Astronomie der Universit\"at Heidelberg, Astronomisches Rechen-Institut, M\"onchhofstr. 12, 69120 Heidelberg, Germany; 
\email{andreas.koch@uni-heidelberg.de}
  \and University of California Los Angeles, Department of Physics \& Astronomy, Los Angeles, CA, USA
  \and Max-Planck-Institut f\"ur Extraterrestrische Physik, Gie\ss enbachstra\ss e 1, 85748 Garching, Germany
\and Institute of Physics, Laboratory of Astrophysics, Ecole Polytechnique F\'ed\'erale de Lausanne (EPFL), 1290 Sauverny, Switzerland 
   }
\date{}
\abstract{
The shapes of galaxies, in particular their outer regions, are important guideposts to their formation and evolution. 
Here we report on the discovery of strongly box-shaped morphologies of the, otherwise well-studied, elliptical and lenticular galaxies NGC 720 and NGC 2768 
from deep imaging. 
The boxiness is strongly manifested in the shape parameter $A_4/a$ of $-0.04$ in both objects, and also 
significant center shifts of the isophotes of $\sim$ 2--4 kpc are seen. 
One  reason for such asymmetries commonly stated in the literature is a merger origin, 
although the number of such cases is still sparse and the exact properties of the individual boxy objects is highly diverse.
Indeed, for NGC 2768, we identify a progenitor candidate { (dubbed {\em Pelops})} 
in the residual images, which appears to be a dwarf satellite that is currently merging with NGC~2768. 
At its absolute magnitude of M$_r$ of $-$12.2 mag, the corresponding Sersic radius of 2.4 kpc is more extended than 
those of typical dwarf galaxies from the literature. However, systematically larger radii are known to occur in systems that are in tidal disruption. 
This finding is bolstered by the presence of a tentative tidal stream feature on
 archival GALEX data. 
Finally, further structures in the fascinating host galaxy comprise
 rich dust lanes and a vestigial X-shaped bulge component. 
}
\keywords{Galaxies: formation --- Galaxies: halos ---  Galaxies: individual: NGC 720, NGC 2768 ---   Galaxies: interactions --- 
Galaxies: elliptical and lenticular, cD --- Galaxies: structure}
\maketitle 
%
%
%
%
\section{Introduction}
The morphology of galaxies holds important clues about their formation and evolutionary processes, be it in the spiral-vs-elliptical dichtomy, prominently 
illustrated in the Hubble tuning fork and its extensions \citep{Hubble1926,deVaucouleurs1959}, or via including irregular objects that often are the result of past { or} 
 ongoing tidal interactions or mergers \citep{Arp1966,Tal2009}. 

One particular case are galaxies with isophotes  that  significantly deviate from smooth ellipses\footnote{ Commonly quantified in terms of the 
 fourth-order Fourier parameter in an isophote analysis 
 The nomenclature of these parameters differs amongst the literature. 
Here, we follow the internal naming of our used IRAF {\em ellipse} task, which denotes the  isophote-intensity weighted
fourth moment ($B_4/a$) as $A_4/a$  \citep[e.g.,][]{BenderMoellenhoff1987,Jedrzejewski1987,Bender1988,Bender1989}
and similar for the third moments ($B_3/a\equiv A_3/a$).
}. 
The most extreme deviations 
can tend towards a ``disky'' ($A_4/a>0$) or a ``boxy''  ($A_4/a<0$) shape.
\citet{Graham2012} report on a rectangular dwarf galaxy (LEDA 074886; $M_R=-17.3$ mag) with a very high negative boxiness parameter ranging from $A_4/a$ = $-0.05$ to $-0.08$ between 3 and 5 kpc, 
which they dubbed the ``Emerald Cut Galaxy'' (in the following ``ECG''). 
One possible reason for the ECG's boxiness discussed in their work is the edge-on merger of two spiral galaxies. However,  
\citet{Graham2012} 
 emphasize that there are only a few highly boxy examples known in the literature, yet the details of their shapes are diverse, and 
accordingly pinning down one single formation channel is unrealistic. This picture has hardly changed in the literature over the past decade. 

Here, we report on the discovery of boxy morphologies in the halos of two, otherwise well-studied, galaxies in the Local Volume: 
NGC 720 (E5) and 2768 (E6/S0), each with stellar masses of a few $\times$10$^{11}$ M$_{\odot}$ \citep{Rembold2005,Forbes2012,Pastorello2014}. 
\citet{Rich2019} lists their halo shape as ``boxy'' and ``round'', respectively, purely based on visual inspection.
Prompted by previous,  shallower works that did not detect any peculiarities in these objects, this begs the 
question of whether their boxiness is an intrinsic property of the individual galaxy or if it might 
 represent the general presence of disks or other substructures \citep[e.g.,][]{Pasquali2007}.
 To this end, we employ new deep imaging from the ``Halos and Environments of Nearby Galaxies'' (HERON) survey \citep{Rich2019},
bolstered by  archival data from the Sloan Digital Sky Survey (SDSS), and the Galaxy Evolution Explorer (GALEX; 
\citealt{Bianchi2000,Morrissey2007}) 
 to investigate the shapes of those two particularly boxy galaxies. 
This paper is organized as follows: in Sect.~2, we describe the images that are the basis of our study. Sect.~3 is dedicated to the structural analysis of 
the two galaxies. 
Finally, we discuss our findings in terms of the formation histories of either object in Sect.~4.
\section{Observations: Centurion 28 imaging}
Out of the sample of 119 HERON galaxies the two objects of this study were chosen by eye based on their optical appearance and indications of boxiness.
In the following, we briefly introduce the two data sets employed in our structural analysis.

The images used in this work were taken as part of the HERON survey
and details of their reduction are given in \citet{Rich2019}, alongside a general characterization of the galaxies, for instance in terms of their halo sizes.
Imaging for the two targets was acquired  in Oct. and Nov. 2011 with 
the 28-inch Centurion (C28) telescope at the Polaris Observatory Association in Lockwood Valley, California \citep{Rich2012,Brosch2015,Koch2017_1661,Rich2019}. 
The pixel scale of the detector is 0.82$\arcsec$ pixel$^{-1}$, which for NGC 720 has been resampled by a factor of two in either dimension.
This corresponds to 218 and 88 pc per pixel at the adopted distances of NGC 720 and 2768, respectively\footnote{The distances were taken as 27.38 Mpc and 22.15 Mpc
to NGC~720 and NGC2768, respectively (see \citealt{Rich2019} and references therein).}.

 The fields around the galaxies were exposed for 13$\times$300 s (NGC 720) and 3$\times$300 s (NGC 2768)
 using a broad-band Astrodon Luminance filter, which has  a bandwidth  from 4000 to 7000 \AA~and thus  effectively acts as a wide Sloan $r$-filter.
As a result, the images reach surface brightnesses of 29.9 and 28.9 mag\,sq.arcsec$^{-1}$ for NGC 720 and NGC 2768, respectively. 
The seeing conditions of the observations were rather low, at 6.4$\arcsec$ for NGC 720
and 3.5$\arcsec$ for NGC 2768. 
Fig.~1 shows the full C28 images for either galaxy.

The fundamental coordinate system was attached to the images using the public service \url{astrometry.net} \citep{Lang2010}, which builds on blind pattern matching.
Finally, we obtained the 
photometric calibration by performing aperture photometry of stellar sources within the images using SExtractor \citep{Bertin1996}, 
and cross-matching the results  to the $r$-band photometry of the 14th data release of the Sloan Digital Sky Survey (SDSS DR14;  
\citealt{Abolfathi2018}).

\begin{figure}[htb]
\centering
\includegraphics[width=1\hsize,angle=180]{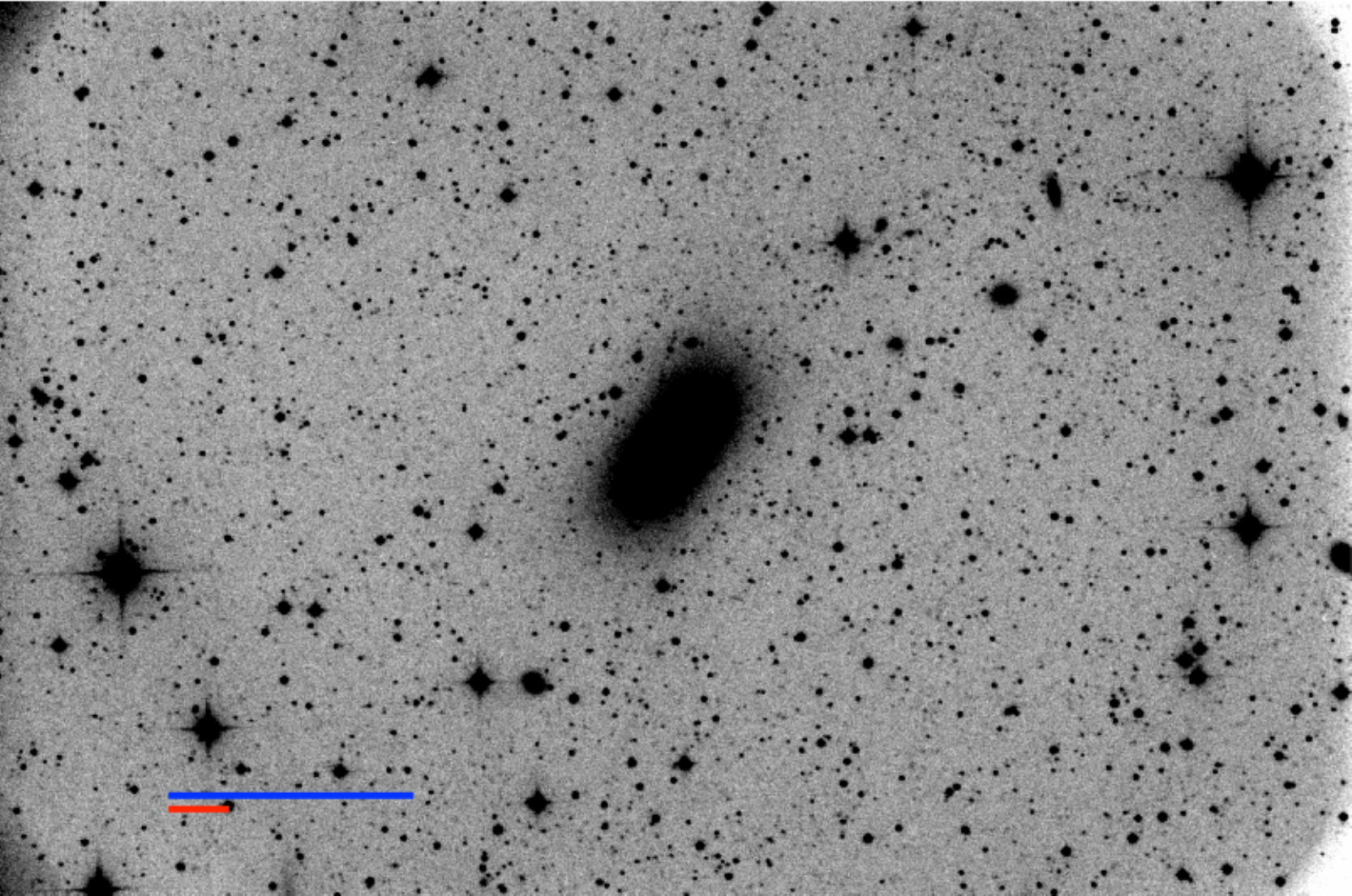}	
\includegraphics[width=1\hsize]{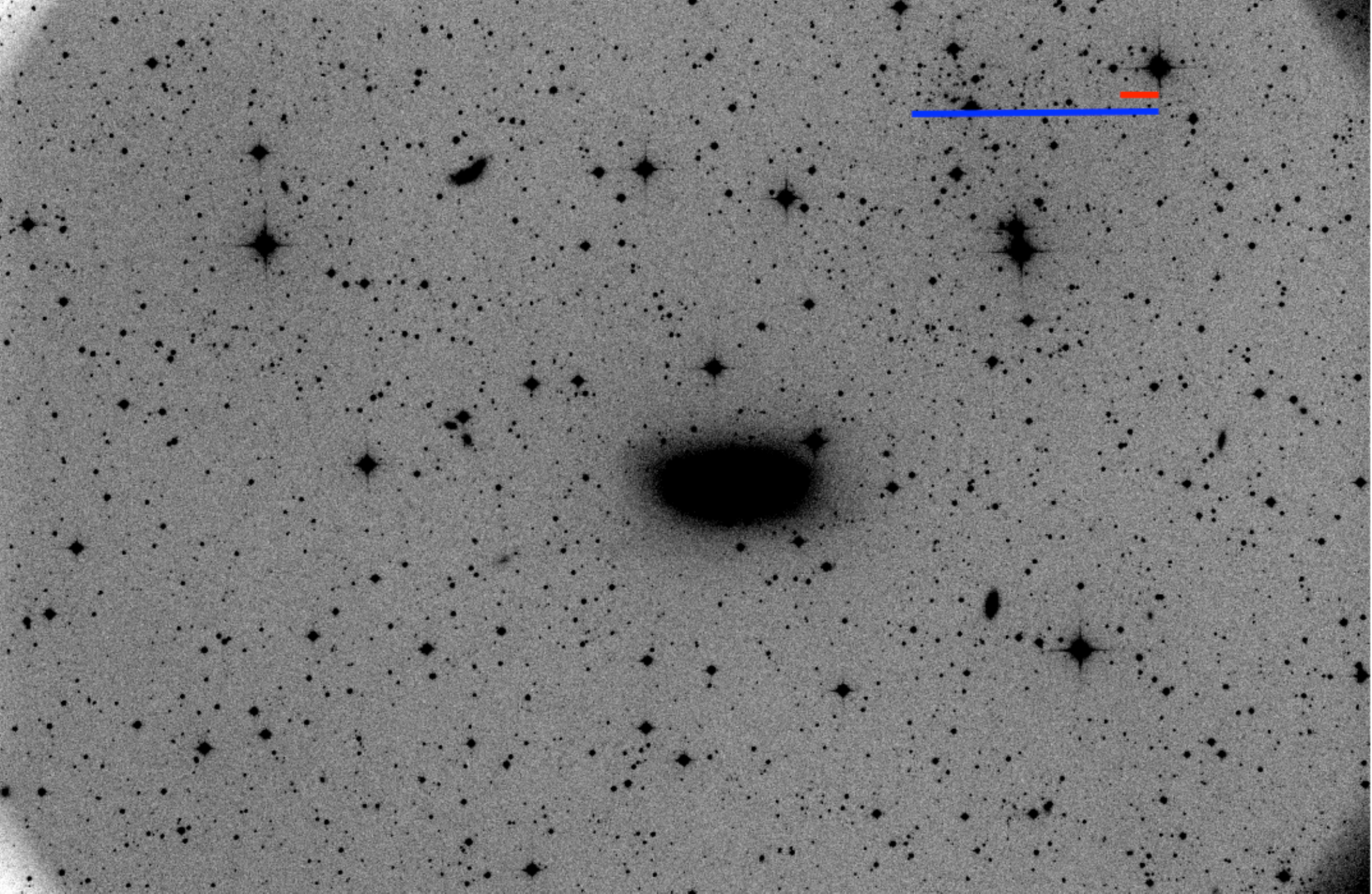}	
\caption{Full C28 images of NGC 720 (top panel) and NGC 2768 (bottom panel) on a linear stretch. 
North is up, East is left. We also indicate scale bars of 10$\arcmin$ (blue line) and 20 kpc (red line) at the adopted distance to the galaxies.
 }
\end{figure}
\section{Analysis}
We performed an isophotal analysis of both galaxies using IRAF's {\em ellipse} task (as was also done by \citealt{Graham2012}), which fits basic ellipse parameters to the flux, with the option 
to include higher order parameters, where our chief focus lies on $A_4/a$. 
To prepare the images, stellar sources were masked within IRAF
and we used a 2$\sigma$-clipping within {\em ellipse} to interpolate over any possible residual flux.
The resulting radial profiles of the key  
shape parameters are shown in Figs.~2 and 4 for either galaxy down to 3$\sigma$ above the sky, and in the following we discuss the implications for their boxiness and potential formation scenarios. 
For the case of the surface brightness profile, the magnitudes have been corrected for extinction by 
{ $A_r=0.036$} mag  for NGC 720 and $A_r=0.103$ mag for NGC 2768 \citep{Schlafly2011}.
\subsection{NGC 720}
In the first series of HERON papers, \citet{Rich2019}  { traced  NGC 720 down  to a surface brightness of 29.9  mag\,arcsec$^{-2}$ and
stated a halo diameter at a flat-rate 28.0  mag\,arcsec$^{-2}$  by ocular inspection. The respective halo ``size''  thus extends to { 23 times its half-light radius (at 6.7 kpc)}, 
and the profile sampled in this work reaches to about 10 half-light radii. 
We note, however, that our profiles are truncated at  a surface brightness of 3$\sigma$ above the background.}
NGC 720 has been classified as an E5 galaxy 
with a total mass of 3.29$\times$10$^{11}$ M$_{\odot}$ \citep{Rembold2005} and only little rotation. 
Also our ellipticity profile (middle left of Fig.~2) meanders around the corresponding value of $\epsilon$$\sim$0.4--0.6.
\begin{figure}[htb]
\centering
\includegraphics[width=1\hsize]{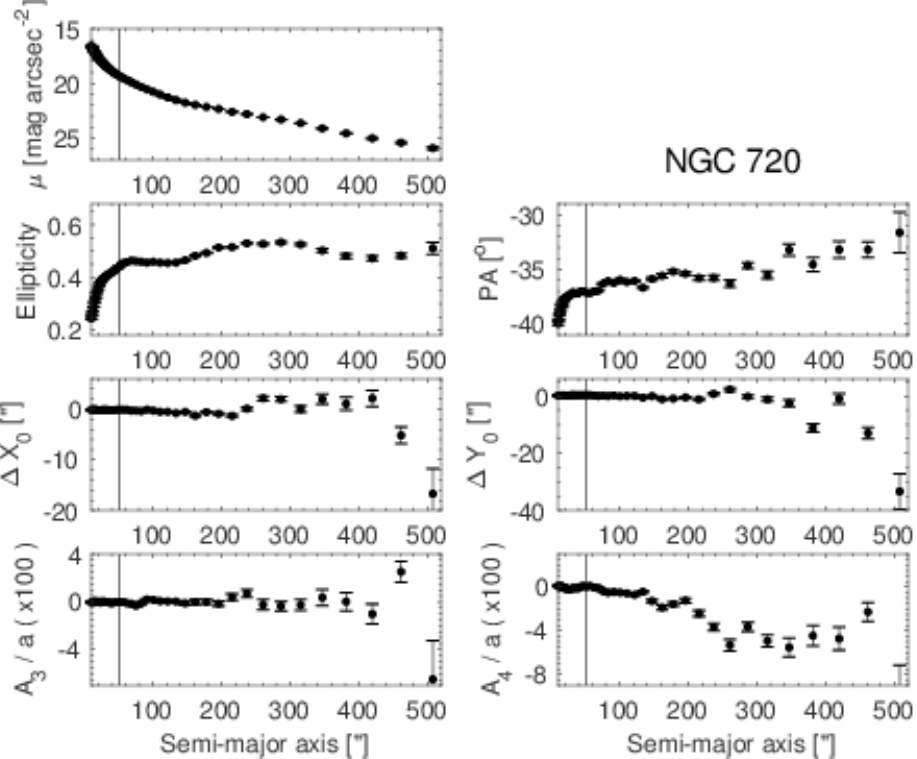}
 \caption{Photometric and morphological parameters of NGC 720 from {\em ellipse}, shown as a function of semi-major axis distance. The half-light radius is indicated.}
\end{figure}

Of main interest for our present work is the intensity-weighted mean boxi-/diskiness parameter $A_4/a$. This is typically measured from 2 seeing radii to 1.5 half-light radii
\citep[e.g.,][]{Carter1978,Carter1987,Bender1989,Hao2006,Graham2012}. 
To guide the eye,  galaxies that are labelled ``boxy'' (excluding dwarfs) have parameters ranging from  $-0.02$ to around zero, 
with values reported { as low as $-$0.04} \citep[e.g.,][]{Hao2006}. 
{ The boxiness parameters around $-$0.04 are effectively rare in the literature.}
Here, the ECG stands out in having very low values of $-$0.05 down to $-0.08$ between 3 and 5 kpc
\citep{Graham2012}.
The boxy nature of NGC 720 has already { been} noted by \citet{Rich2019} ``by eye'' and is now quantitatively confirmed (bottom left panel of Fig.~2), reaching an $A_4/a$ of $-0.04$ in its outskirts, which uniquely classifies this galaxy as a boxy one, 
albeit to a lesser extent than the ECG.

Moreover, we find a strongly varying center position for NGC 720 throughout the annuli, varying by 10--20 px ($\sim$2--4 kpc at the adopted distance). { This} is likely due
to the same event that caused the boxiness of the isophotes, which we conjecture to be a merger (see also Sect.~3.2.2).   
The residual image (Fig.~3) displays further butterfly-shaped features, which are a common feature  
if the disk component is not properly modelled and removed, thus revealing complexities in the disk such as dust lanes. 
\begin{figure}[htb]
\centering
\includegraphics[width=1\hsize,angle=180]{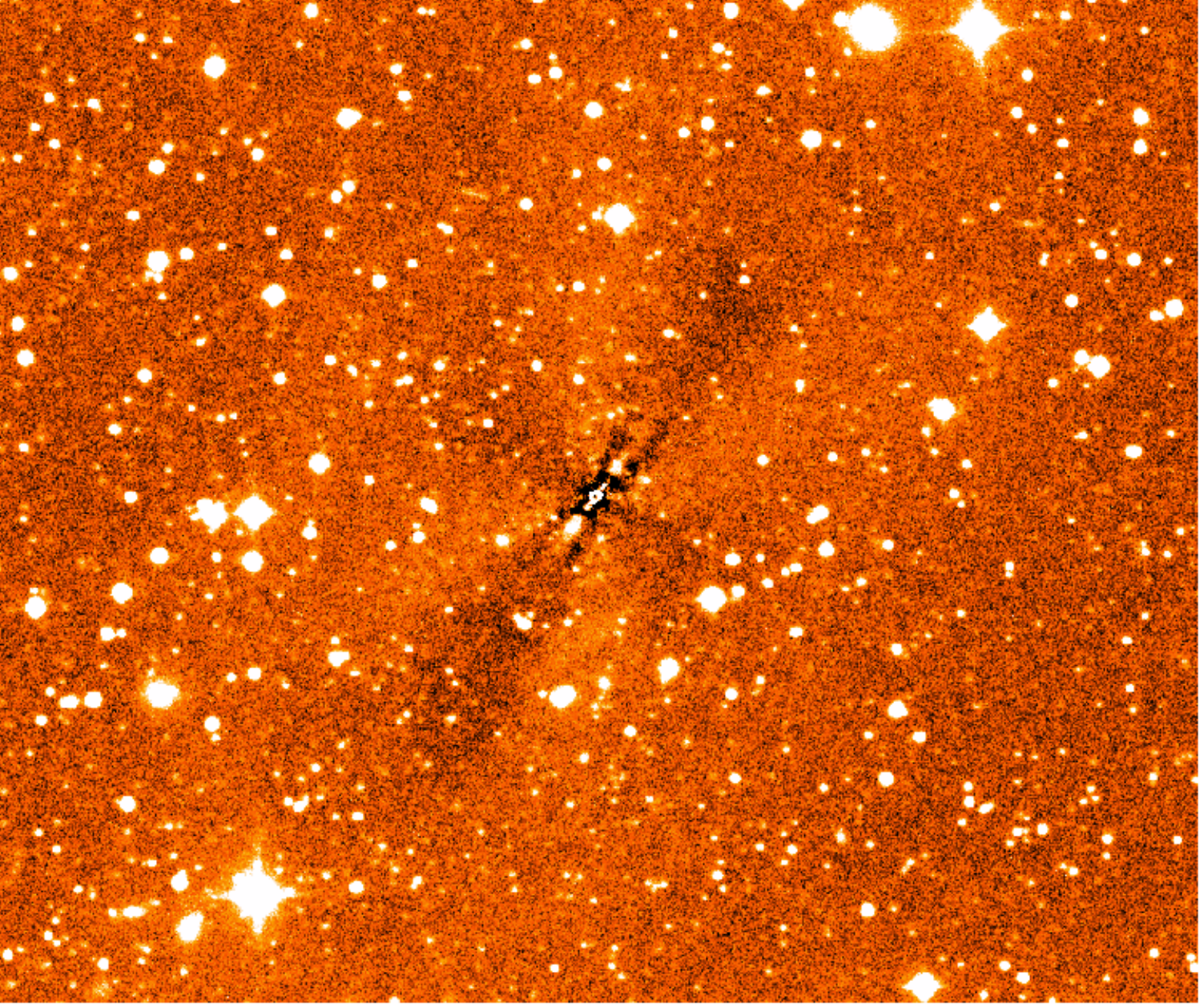}
\caption{Model subtracted image of NGC~720, covering  $25\arcmin\times 20\arcmin$. North is up, East is left.}
\end{figure}
\subsection{NGC 2768}
{ This galaxy has been traced down to 28.9 mag\,arcsec$^{-2}$ by \citet{Rich2019}
and its diameter at the 28 mag\,arcsec$^{-2}$ level is reported as 96 kpc, 
 corresponding  to $\sim$13 effective  radii (the latter being 7.6 kpc)}. In turn, the data under scrutiny here
cover approximately six half-light radii (above 3$\sigma$ of the sky) before the background hampered a further meaningful analysis. 
NGC 2768 already appears boxy to the eye (Fig.~1), which is bolstered by the shape profiles in Fig.~4 (bottom right panel). The low values of $A_4/a$ of $-0.04$
in its outer regions render it a clear contender for a boxy galaxy. Its nature as a purported lenticular to elliptical (E5) galaxy is also sustained by our derived ellipticity profile (middle left panel of Fig.~4). 
\begin{figure}[htb]
\centering
\includegraphics[width=1\hsize]{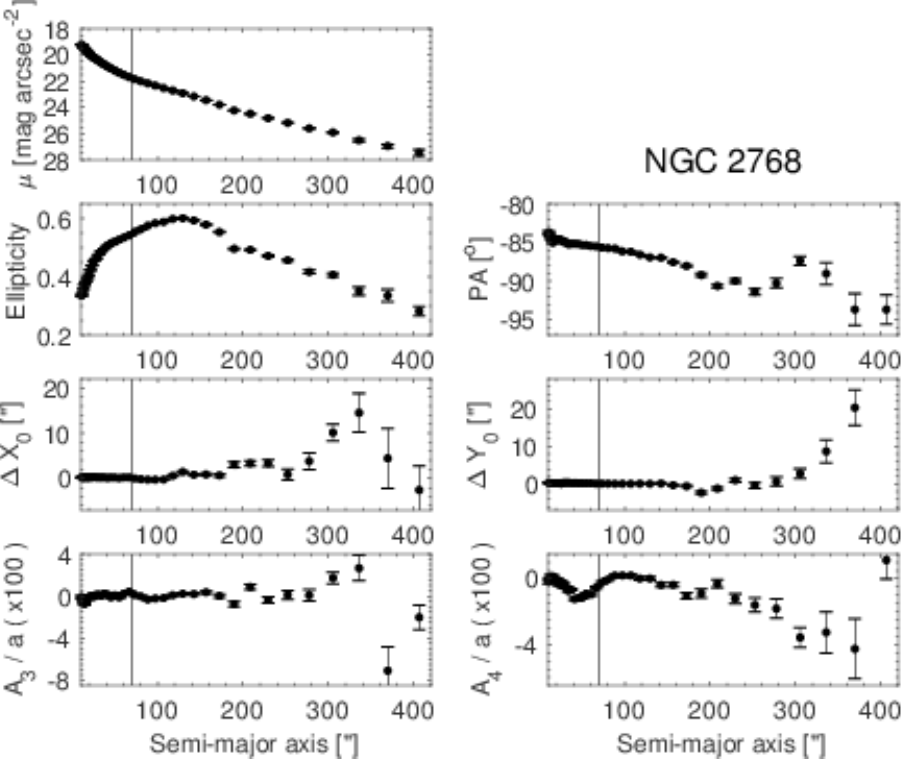}
 \caption{Same as Fig.~2, but for NGC~2768.}
\end{figure}
As for NGC 720, we find a significant center shift across the full isophotes, amounting to as much as 20 px in $x$ and $y$-direction, which corresponds to 
$\sim$1.8 kpc at the used pixel scale and adopted distance to the target galaxy. 
{  In fact, the most pronounced  shifts for both galaxies, appears after 200--300 arcsec}.

A hint of the peculiarity in this galaxy was already found by \citet{Pulsoni2018}, who measured the kinematics of planetary nebulae out to 5 R$_e$, resulting in a non-point-symmetric distribution.
They found this object to be a 
 fast rotator out to large radii,  and they quantified its asymmetry with similar parameters to ours (viz., $c_4$ and $s_4$, accounting for sine and cosine projections). 
 \citet{Pulsoni2018} judged these asymmetries as ``likely real'' as they have also  already been seen in the deep optical images of \citet{Duc2015}.
\subsubsection{A disk in NGC 2768}
A prominent dust lane, as visible as black stripes on the model-subtracted image of NGC 2768 (Fig.~5), had already been noted by \citet{Kim1989} and a hint of it 
can also { be } seen on our original C28 image (Fig.~1) when using a proper stretch. Moreover, we note the possible presence of  a 
vestigial x-shaped bulge structure as is also known to exist in the Milky Way \citep{McWilliam2010}. 
\begin{figure}[htb]
\centering
\includegraphics[width=1\hsize,angle=180]{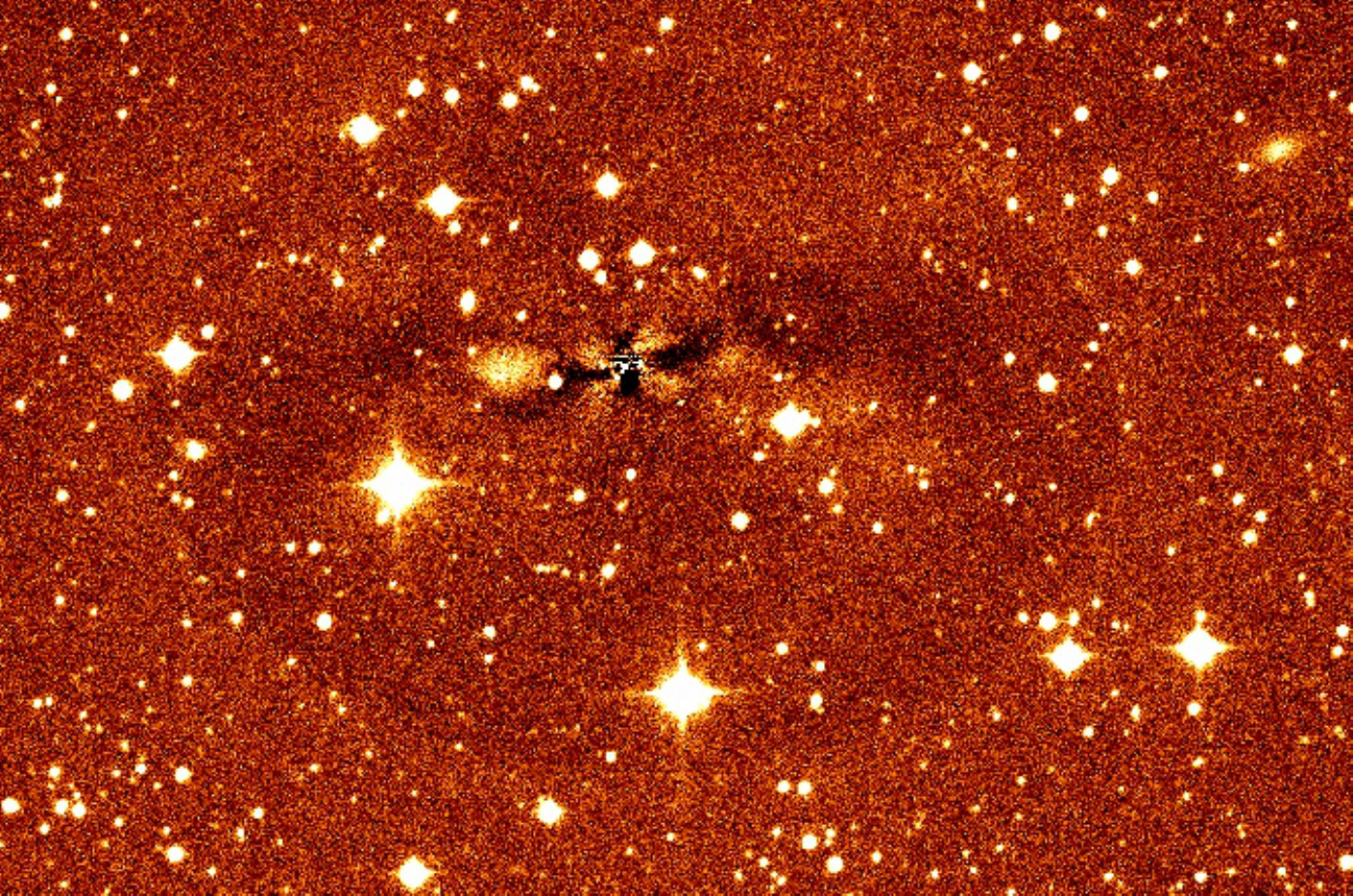}
\caption{Model subtracted image of NGC~2768, covering  $20\arcmin\times 13\arcmin$.}
\end{figure}
An identical isophotal analysis on the HST data as on our C28 imaging reveals the exact same features.
Finally, we note that we have also investigated archival SDSS images in the $r$-band, which confirm the presence of the 
dust. 
\subsubsection{The progenitor that built NGC 2768}
Fig.~6 is the result of masking not only the stars before running {\em ellipse}, but also masking the dust features mentioned in Sect.~3.2.1.
This model subtracted image clearly shows the presence of a large plume towards the West of NGC 2768's center, 
which we consider to be the 
ongoing merger that caused the strong distortions of NGC~2768's isophotes, { which we henceforth dub  {\em Pelops}}\footnote{ Son of Tantalos. According to
Greek mythology, Pelops was ``tidally disrupted'' (rather, chopped to pieces) to feast the Gods.}.
Using {\em GALFIT} \citep{Peng2002} we fitted a Sersic-profile to the model-subtracted image. 
The result of this fit is shown in { Fig.~6} and its basic parameters are summarized in Table~1. 
{ {\em GALFIT} explicitly accounts for the point spread function (PSF) of the images in its fitting so that the stated radii are the ones obtained after decovolution 
with the PSF profile. Similarly, this
is considered (internally within {\em GALFIT}) for the error analysis.}
\begin{figure*}[htb]
\centering
\includegraphics[width=0.3\hsize]{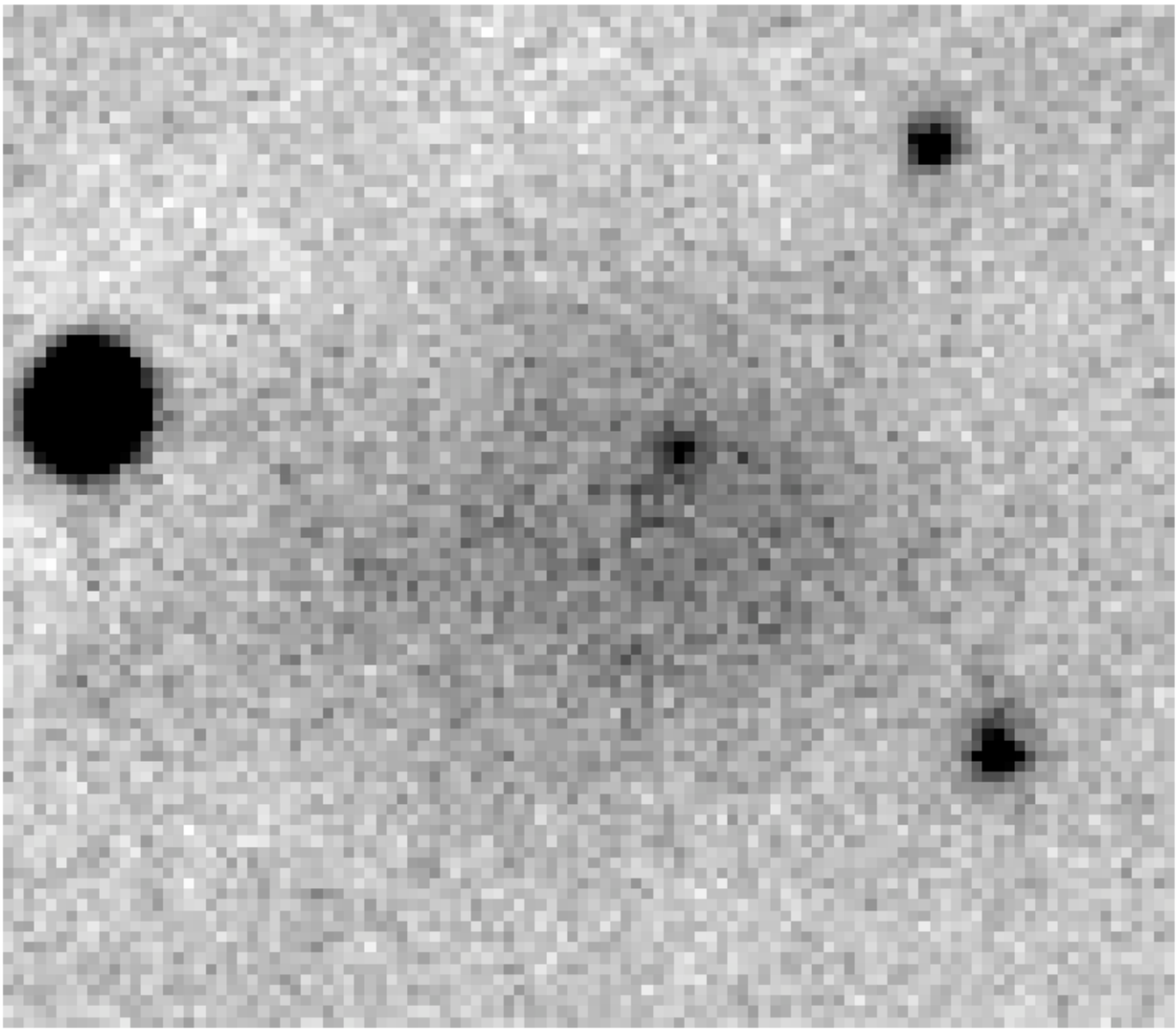}	
\includegraphics[width=0.3\hsize]{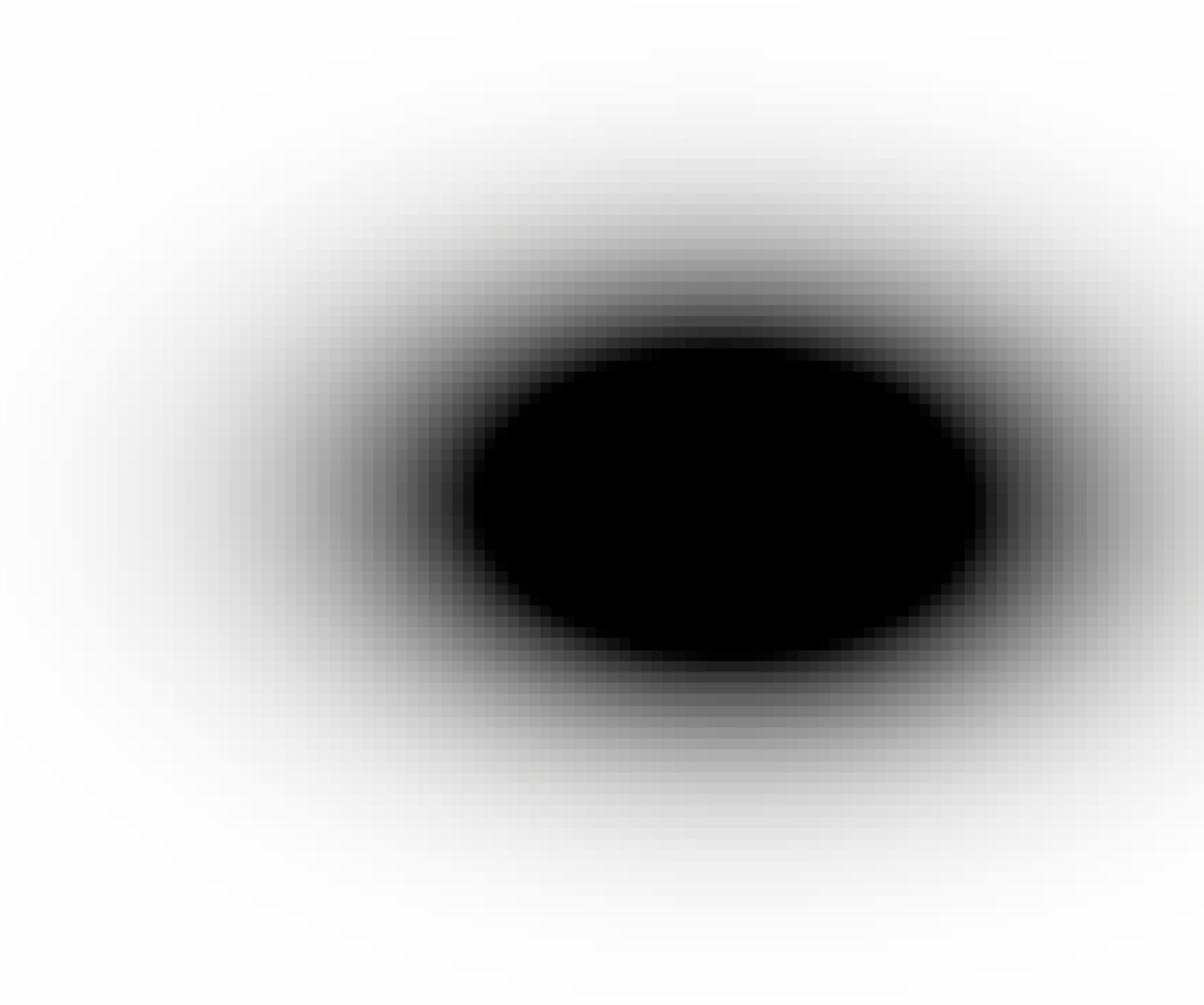}	
\includegraphics[width=0.3\hsize]{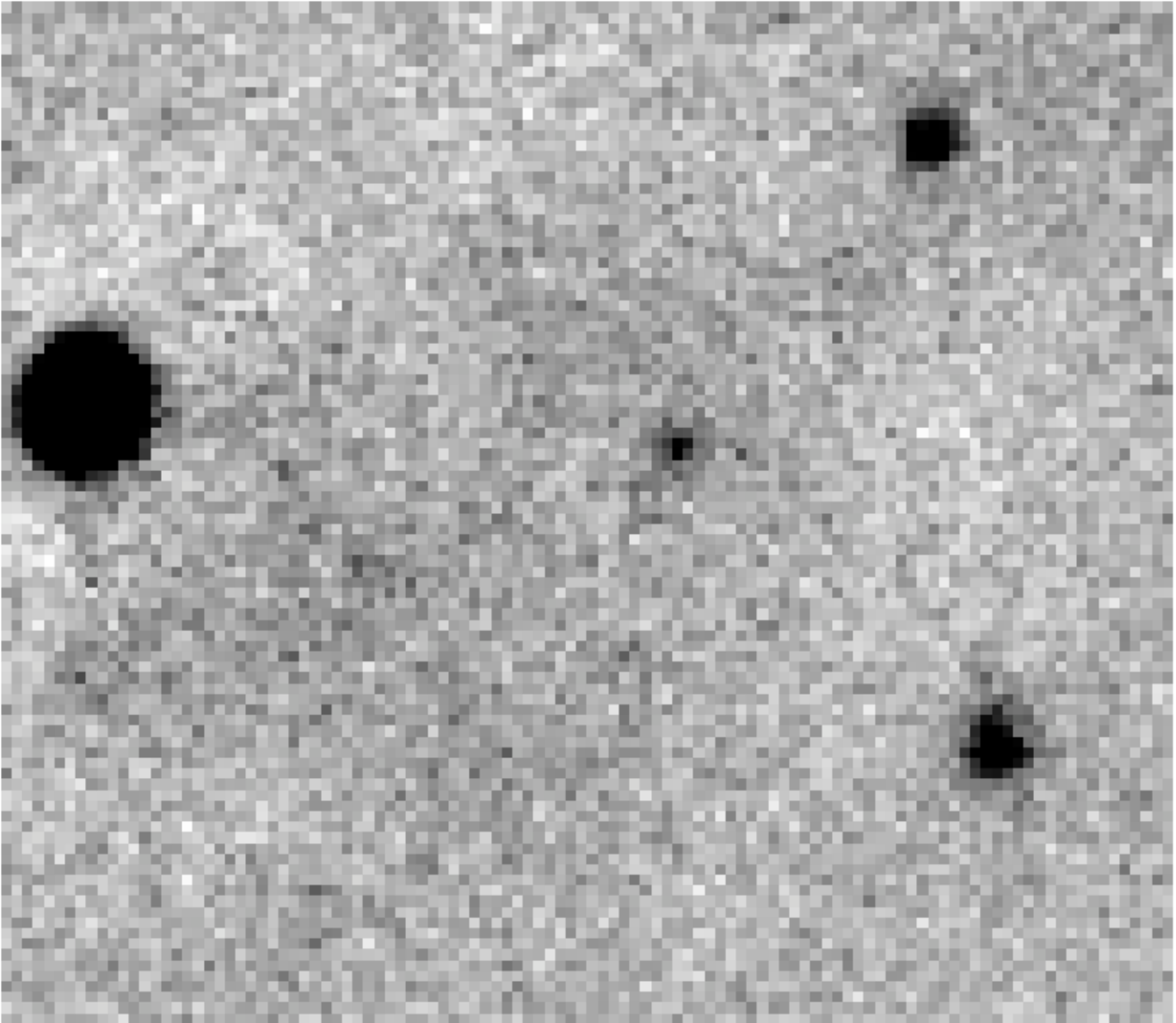}	
\caption{Left panel: a $1.5\arcmin\times 1.3\arcmin$ image  (thus encompassing $\sim$4 half-light radii) centered on { \em Pelops}. The middle panel shows the best-fit ($\chi^2/\nu$=0.96)
{\em GALFIT}-model, and the right panel depicts
the residual image.}
\end{figure*}
\begin{table}[htb]
\caption{Parameters of the NGC 2768 merger candidate, { \em Pelops}.}             
\centering          
\begin{tabular}{ccc}     
\hline\hline       
Parameter & Value & Method\tablefootmark{a} \\
\hline
$\alpha$ & 09:11:24.6 & G, C28\\
$\delta$ &$+$60:02:18.2 & G, C28 \\
M$_R$  & $-$12.2 $\pm$ 0.2 & G, C28 \\
$\mu_e$ & 22.96 $\pm$ 0.01 & G, C28\\
$r_e$ & 2.4 $\pm$ 0.3 kpc &  G, C28 \\
$n$ & 0.37 $\pm$ 0.02& G, C28 \\
$e$ & 0.55 $\pm$ 0.01& G, C28 \\
P.A. & 88.4$^{\circ}$ $\pm$ 1.1$^{\circ}$ & G, C28 \\
$u_0$ & 19.19 $\pm$ 0.26 & A, SDSS \\
$g_0$ & 17.93 $\pm$ 0.14 & A, SDSS \\
$r _0$ & 17.00 $\pm$ 0.09 & A, SDSS \\
$i _0$ & 16.65 $\pm$ 0.08 & A, SDSS \\
$z_0$ & 16.50 $\pm$ 0.07 & A, SDSS \\
($g-r$)$_0$ & 0.93$\pm$0.17 & D \\
($r-i$)$_0$ & 0.35$\pm$0.12 & D \\
\hline
\hline                  
\end{tabular}
\tablefoot{
\tablefoottext{a}{G, C28: {\em GALFIT} values from our C28 images; A, SDSS: Aperture photometry within $r_e$ on the SDSS residual images; D: Derived}
}
\end{table}

Assuming this blob feature \citep{Casey2023} to be at the same distance as NGC 2768, we determine its absolute magnitude as M$_r=-12.2\pm$0.12 mag.
The best-fit Sersic index was determined as 0.37$\pm$0.02,  which is
 rather small and  typical of disrupting galaxies and indicates that a simple Sersic-modeling
is not adequate anymore. 
Furthermore, we determine an axis ratio of 0.55$\pm$0.01 and the position angle of { \em Pelops} of 88.4$^{\circ}\pm1.09^{\circ}$. 
\begin{figure}[htb]
\centering
\includegraphics[width=1\hsize]{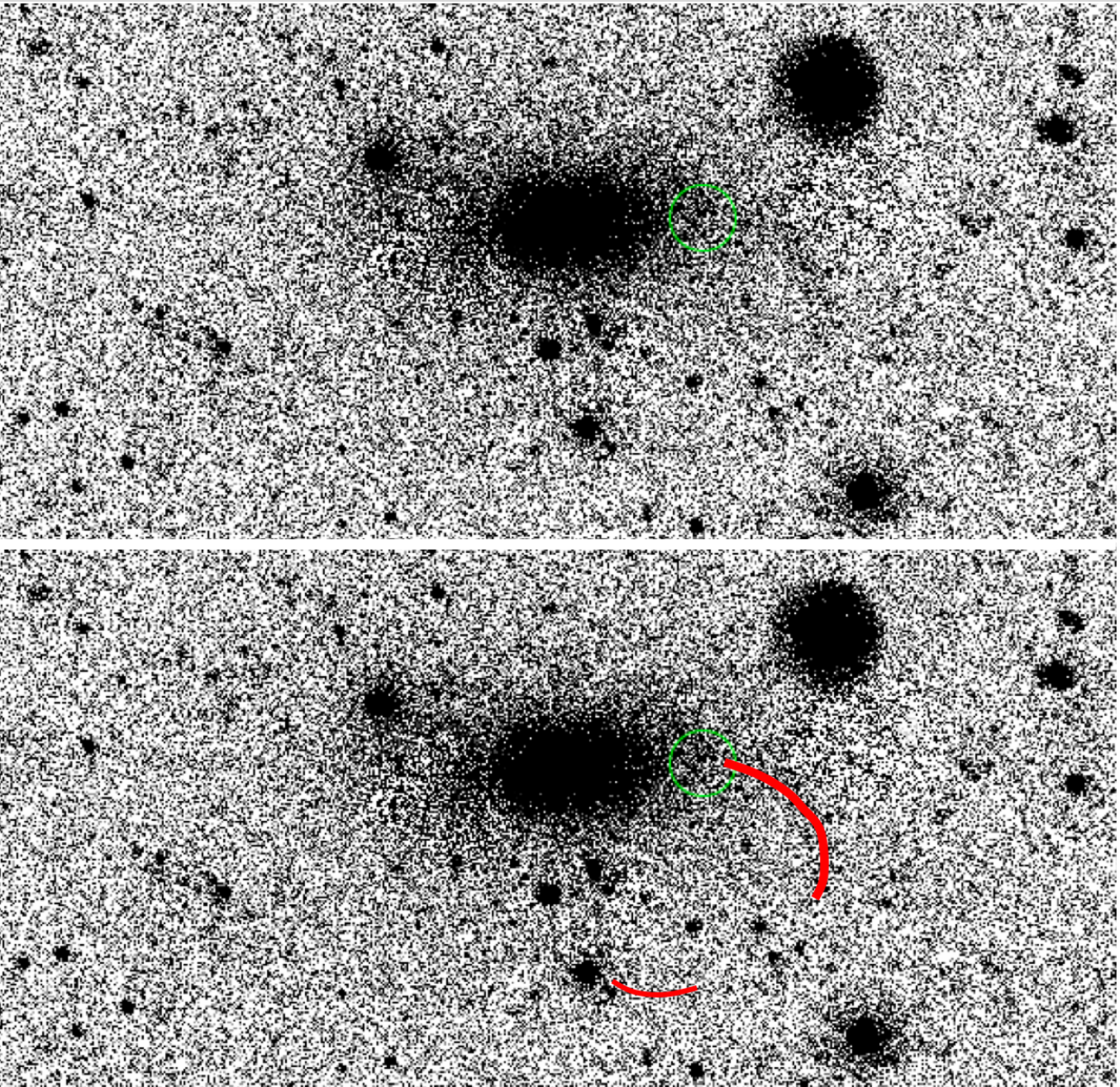}	
\caption{GALEX NUV image of NGC 2768. The location of the { {\em Pelops} overdensity} identified in Fig.~6 is indicated with a green circle.
The purported stream is highlighted by the red lines in the bottom panel. The image covers $14\arcmin\times7\arcmin$.}
\end{figure}
To find further evidence of the reality of the merging galaxy, we consulted archival data from the Galaxy Evolution Explorer (GALEX; \citealt{Bianchi2000,Morrissey2007}), which targeted the surroundings of NGC~2768
in the context of the NGA campaign \citep{GilDePaz2007} in Feb. 2005. While no detection can be made in the Far-ultraviolet (UV) image, the 182-s exposure in the Near-UV
clearly depicts the host galaxy (Fig.~7). 
Furthermore, we note the presence of a vestigial stream and possibly an extension into a further arc to the South of the host galaxy.
These are indicated by eye in the bottom panel of Fig.~7.
A natural suspicion is that the structure we see { could be a} reflection. However, according to the GALEX documentation\footnote{\url{http://www.galex.caltech.edu/wiki/Public:Documentation/Chapter_8#Ghosts}}, ghosts chiefly appear at 30--60$\arcsec$ above and below the bright source 
along the $y$-direction on the detector. In our case, the arcs appear at a much larger separation of 140$\arcsec$ with an even larger extent and a shape that does not 
resemble the ghostly donut-shapes. Therefore, we deem it unlikely, that the purported stream is an artefact. 
\subsubsection{Properties of { \em Pelops} from SDSS images}
The same feature also stands out in identically model-subtracted SDSS images in the $g$, $r$, and $i$-bands, confirming that we are most likely seen a real feature, while being
fainter in the $u$- and $z$-bands. The (lack of) depth of the SDSS prevents us from obtaining any meaningful structural or photometric parameters from {\em GALFIT}.
However, we performed {\em aperture} photometry by simply adding the calibrated flux on each image within one effective radius, both of the merger candidate and of the host galaxy. 
The magnitudes were dereddened using the dust maps of \citet{Schlafly2011} and the extinction law of \citet{Cardelli1989}.
Here, it worth noticing that both the $g-r$ and $r-i$ colours of both objects are in very good agreement to within the (Poisson) errors.  

Based on our photometry, we consulted the E-MILES simple stellar population (SSP) models \citep{Vazdekis2016}, which we 
computed for a LMC-like metallicity and with the universal initial mass function of \citet{Kroupa2001}
for 53 ages between 0.03 Gyr and  14 Gyr.
For each age, we
 varied the {\em intrinsic} reddening, A$_V$, between 0 and 2 in steps of 0.05.  
Next, we computed the reduced $\chi^2_{\nu}$ between the predicted SSP colours and the ones measured from the SDSS images. 
This results in a best-fit age of $6.5^{+5.5}_{-3.8}$ Gyr and an intrinsic reddening of $0.15\pm0.15$ mag. 
The according mass-to-light (M/L) ratio  in the $r$-band is found to be 1.9$^{+1.0}_{-0.9}$ in Solar units, which is rather on the low side for a typical dwarf galaxy \citep[e.g.,][]{Koch2009Review}. 
This would imply, adopting the satellite's absolute magnitude, 
a total mass of $\sim$10$^7$ M$_{\odot}$.  
\subsubsection{{ \em Pelops} in context}
The surface brightness at the effective radius is fully in agreement with those of dwarf galaxies in various environments
(Fig.~8, middle panel). However, 
at its absolute magnitude this merging candidate appears too large by a factor of a few, when compared to
typical galaxies of similar magnitude.  
Its corresponding half-light radius is 2.4$\pm$0.3 kpc, and an  
investigation of systematically more extended objects is often used to confirm the presence and absence of tidal disruptions
\citep{Koch2017_1661}. Indeed, the contender within NGC 2768 lies within 
$\pm$1.5 mag of the strongly disrupted NGC 4449B \citep{Rich2012} and HCC-087 \citep{Koch2012HCC}
at similarly large radii. 
We note, however, that the { shown literature sample  is given} in the Johnson-Cousins R-band (and in parts transformed from Sloan $g$- and $i$-band
magnitudes; \citealt{Byun2020}), whereas 
our value is in Sloan-$r$ and converted from our luminance measurement so that a slight offset in magnitude can be inherent.
We further note that we adopted a single Sersic profile, which does not account for tidal features, 
while other galaxies 
in the literature may employ other types of profiles, adding to the discrepancy.
Two data points in the bottom panel Fig.~8 are worth mentioning: firstly, the Local Group dwarf spheroidal And~XIX, which, at M$_{\rm R}=-9.74$ and R$_{\rm h}=1.7$ kpc \citep{McConnachie2012}, stands out 
in the radius-magnitude diagram. Also this ultra-diffuse object is a result of tidal interactions with its host galaxy, M31 \citep{Collins2020}. 
Secondly, Antlia 2 has been named ``an enormous Galactic dwarf satellite''  \citep{Torrealba2019}, the properties of which are also indicative of a strong tidal 
evolution.
We therefore conclude that the overdensity is, by effective and absolute magnitude, a common dwarf galaxy in the process of tidal interactions with the host, NGC~2768.
\begin{figure}[htb]
\centering
\includegraphics[width=1.4\hsize]{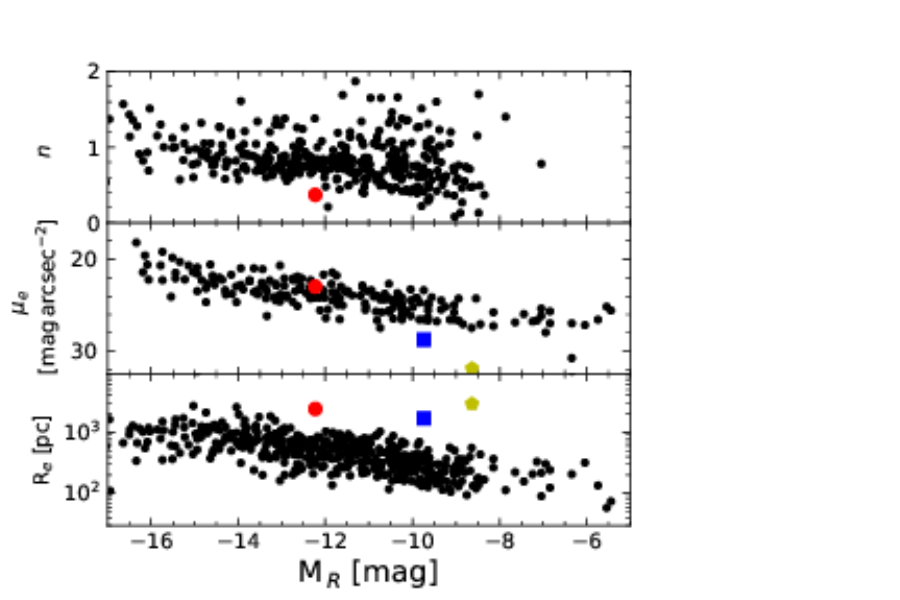}	
\caption{Location of { \em Pelops} (red point) on the magnitude-radius plot (bottom panel) and in relation with the surface brightness at effective radius (middle panel)
and the Sersic-index (top panel). Data from dwarf galaxies in 
various groups and clusters are shown as black points 
\citep{Chiboucas2009,Mueller2015,McConnachie2012,Munoz2015,Park2017,Byun2020}. 
Indicated in blue and green are the dwarf spheroidals And~XIX and Antlia 2, which are the most extended objects in the Local Group.} 
\end{figure}

To place the merging galaxy in context, we compared its absolute magnitude 
with the Local Group dwarf sample \citep{McConnachie2012}. Albeit given in the V-band, the satellite to { NGC~2768} appears similar
to Andromeda II, which, intriguingly, might be the remnant of a merger in itself \citep{Amorisco2014}. 
Our candidate has a luminosity of $\sim$5$\times10^6$\,L$_{\odot}$. 
Adopting the M/L-ratio determined above, this results in a mass of the dwarf candidate on the order of 10$^7$\,M$_{\odot}$.
If the NGC~2768 merger and And~II were of similar 
 type and the present object had the same mass-to-light-ration as And~II ($\sim$20 \citealt{Cote1999}) 
this would yield a mass ten times larger, $\sim$10$^8$\,M$_{\odot}$.
This compares to the host galaxy's (disk plus bulge) mass of $\sim10^{11}$\,M$_{\odot}$ \citep{Forbes2012}. 
No matter which M/L is used, 
we are thus facing a minor merger, which is still seemingly capable of deforming the structure of the galaxy 
and inducing many of its morphological and dynamical properties. 
\section{Discussion}
Upon visual inspection of the rich HERON dataset, we detected unusually boxy isophotes of two, otherwise well-studied, galaxies. 
Such objects are hitherto rare, and in one case we could even identify a merger candidate, which we believe has caused the isophotal 
distortions. 
\citet{Graham2012} suggested an edge-on merger of two disk galaxies as the origin of the ECG, one of the most rectangular galaxies known.
In their scenario, initial gas was driven inwards and formed an inner disk, while larger radii experienced a dissipationless merger. 
The ``boxiness'' parameters of the two well-known objects measured in our study are not as extreme as for the ECG ($A_4/a$ of $-0.08$ vs. our $-0.04$). 
While no remnant could be identified for NGC~720, a 
merger origin of its morphology is bolstered by \citet{Rembold2005}, who identified this galaxy as an unequal-mass merger remnant based on its kinematics from 
longslit spectroscopy. 

NGC 2768 has been classified as an E5 galaxy \citep{Pastorello2014}, but has also been named E6/S0 \citep{Zanatta2018,Rich2019}.
Already early on, it was found that its gas has different kinematics from the stars in the inner regions \citep{Fried1994}, hinting at
a dynamically special history. 
Similarly, \citet{Forbes2012} performed a bulge/disk decomposition based on various photometric and kinematic tracers
(planetary nebulae, stars, and globular clusters) and found that the disk of NGC~2768 rotates rapidly, with its velocity dispersion decreasing with radius.
In contrast, the bulge turned out to be pressure supported with only a slow rotation. 
As the 
resulting ratio of the disk's rotational velocity to its velocity dispersion resembles that
 of a spiral galaxy, \citet{Forbes2012} conclude that NGC~2768 is a transformed late-type galaxy. 
Similarly, \citet{Zanatta2018}, also using globular clusters and planetary nebulae as tracers, note that NGC~2768's red (i.e., old) globular cluster system displays rotation, most 
pronounced at inner radi (R$<$1 kpc), indicating that mergers seem to have played an important role. 
{ Overall, lenticular galaxies} can evolve from spiral galaxies via various processes that remove most of their gas and erase spiral structures 
\citep[e.g.,][]{Byrd1990,Bournaud2005,Zanatta2018}. 
Interestingly, also the 
ECG shows a disk-bar like structure in its very center, where a solid body rotation indicates the presence of a central disk \citep{Forbes2011LEDA,Graham2012}.
An obvious question is how frequent galaxies with such strongly boxy morphologies are in the (Local) Universe.
Several similar contenders are reported in the literature (see \citealt{Graham2012} and references therein), although none of them
display such a boxiness as the ECG. 
For instance, \citet{Bidaran2020} find boxy isophotes in the Virgo cluster dE galaxy VCC 0608, which also has a severe misalignment between the 
photometric and kinematic position angles, indicative of a past merger. 
As \citet{Graham2012} discuss, all boxy galaxies have very individually different properties. 
Here, different galaxy types are covered, and head-tail structures and warps have also been identified.
As a result, it remains difficult to uniquely identify
one tailor-made 
mechanism to produce boxy isophotes. 

\citet{Hao2006} reported that  only 19 out of 847
(i.e., 2.2\%) in their sample of nearby early-type galaxies (elliptical and lenticular) from the SDSS
 are boxy, with $-0.02 < A_4/a < -0.01$,  while the remaining 97.8\% show disky isophotes, i.e. $A_4/a> -0.01$). 
This fraction is in agreement with the number of boxy early-type galaxies in the Virgo Cluster Catalog \citep[][VCC]{Binggeli1985} as identified 
 in \citet{Graham2012} and \citet{Bidaran2020}, which add up to $\sim3$\% the VCC's population.
This highlights that boxy galaxies are still a rare species that await further detections and require more in-depth investigations.
\begin{acknowledgements}
We thank the anonymous referee for a constructive report. The authors warmly thank W. Byun for sharing his data on dwarf galaxies. 
AJKH and AP gratefully acknowledge funding by the Deutsche Forschungsgemeinschaft (DFG, 
German Research Foundation) -- Project-ID 138713538 -- SFB 881 (``The Milky Way System''), subprojects A03, A05, A11, B05. 
OM is grateful to the Swiss National Science Foundation for financial support under the grant number PZ00P2\_202104.
\end{acknowledgements}
\bibliographystyle{aa} 
\bibliography{ms} 

\begin{thebibliography}{55}
\expandafter\ifx\csname natexlab\endcsname\relax\def\natexlab#1{#1}\fi

\bibitem[{{Abolfathi} {et~al.}(2018){Abolfathi}, {Aguado}, {Aguilar}, {Allende
  Prieto}, {Almeida}, {Tasnim Ananna}, {Anders}, {Anderson}, {Andrews},
  {Anguiano}, \& et~al.}]{Abolfathi2018}
{Abolfathi}, B., {Aguado}, D.~S., {Aguilar}, G., {et~al.} 2018, \apjs, 235, 42

\bibitem[{{Amorisco} {et~al.}(2014){Amorisco}, {Evans}, \& {van de
  Ven}}]{Amorisco2014}
{Amorisco}, N.~C., {Evans}, N.~W., \& {van de Ven}, G. 2014, \nat, 507, 335

\bibitem[{{Arp}(1966)}]{Arp1966}
{Arp}, H. 1966, \apjs, 14, 1

\bibitem[{{Bender} {et~al.}(1988){Bender}, {Doebereiner}, \&
  {Moellenhoff}}]{Bender1988}
{Bender}, R., {Doebereiner}, S., \& {Moellenhoff}, C. 1988, \aaps, 74, 385

\bibitem[{{Bender} \& {Moellenhoff}(1987)}]{BenderMoellenhoff1987}
{Bender}, R. \& {Moellenhoff}, C. 1987, \aap, 177, 71

\bibitem[{{Bender} {et~al.}(1989){Bender}, {Surma}, {Doebereiner},
  {Moellenhoff}, \& {Madejsky}}]{Bender1989}
{Bender}, R., {Surma}, P., {Doebereiner}, S., {Moellenhoff}, C., \& {Madejsky},
  R. 1989, \aap, 217, 35

\bibitem[{{Bertin} \& {Arnouts}(1996)}]{Bertin1996}
{Bertin}, E. \& {Arnouts}, S. 1996, \aaps, 117, 393

\bibitem[{{Bianchi} \& {GALEX Team}(2000)}]{Bianchi2000}
{Bianchi}, L. \& {GALEX Team}. 2000, \memsai, 71, 1117

\bibitem[{{Bidaran} {et~al.}(2020){Bidaran}, {Pasquali}, {Lisker}, {Coccato},
  {Falc{\'o}n-Barroso}, {van de Ven}, {Peletier}, {Emsellem}, {Grebel}, {La
  Barbera}, {Janz}, {Sybilska}, {Vijayaraghavan}, {Gallagher}, \&
  {Gadotti}}]{Bidaran2020}
{Bidaran}, B., {Pasquali}, A., {Lisker}, T., {et~al.} 2020, \mnras, 497, 1904

\bibitem[{{Binggeli} {et~al.}(1985){Binggeli}, {Sandage}, \&
  {Tammann}}]{Binggeli1985}
{Binggeli}, B., {Sandage}, A., \& {Tammann}, G.~A. 1985, \aj, 90, 1681

\bibitem[{{Bournaud} {et~al.}(2005){Bournaud}, {Jog}, \&
  {Combes}}]{Bournaud2005}
{Bournaud}, F., {Jog}, C.~J., \& {Combes}, F. 2005, \aap, 437, 69

\bibitem[{{Brosch} {et~al.}(2015){Brosch}, {Kaspi}, {Niv}, \&
  {Manulis}}]{Brosch2015}
{Brosch}, N., {Kaspi}, S., {Niv}, S., \& {Manulis}, I. 2015, \apss, 359, 9

\bibitem[{{Byrd} \& {Valtonen}(1990)}]{Byrd1990}
{Byrd}, G. \& {Valtonen}, M. 1990, \apj, 350, 89

\bibitem[{{Byun} {et~al.}(2020){Byun}, {Sheen}, {Park}, {Ho}, {Lee}, {Kim},
  {Jeong}, {Park}, {Seon}, {Lee}, {Lee}, {Cha}, {Ko}, \& {Kim}}]{Byun2020}
{Byun}, W., {Sheen}, Y.-K., {Park}, H.~S., {et~al.} 2020, \apj, 891, 18

\bibitem[{{Cardelli} {et~al.}(1989){Cardelli}, {Clayton}, \&
  {Mathis}}]{Cardelli1989}
{Cardelli}, J.~A., {Clayton}, G.~C., \& {Mathis}, J.~S. 1989, \apj, 345, 245

\bibitem[{{Carter}(1978)}]{Carter1978}
{Carter}, D. 1978, \mnras, 182, 797

\bibitem[{{Carter}(1987)}]{Carter1987}
{Carter}, D. 1987, \apj, 312, 514

\bibitem[{{Casey} {et~al.}(2023){Casey}, {Greco}, {Peter}, \&
  {Davis}}]{Casey2023}
{Casey}, K.~J., {Greco}, J.~P., {Peter}, A. H.~G., \& {Davis}, A.~B. 2023,
  \mnras

\bibitem[{{Chiboucas} {et~al.}(2009){Chiboucas}, {Karachentsev}, \&
  {Tully}}]{Chiboucas2009}
{Chiboucas}, K., {Karachentsev}, I.~D., \& {Tully}, R.~B. 2009, \aj, 137, 3009

\bibitem[{{Collins} {et~al.}(2020){Collins}, {Tollerud}, {Rich}, {Ibata},
  {Martin}, {Chapman}, {Gilbert}, \& {Preston}}]{Collins2020}
{Collins}, M. L.~M., {Tollerud}, E.~J., {Rich}, R.~M., {et~al.} 2020, \mnras,
  491, 3496

\bibitem[{{C{\^o}t{\'e}} {et~al.}(1999){C{\^o}t{\'e}}, {Mateo}, {Olszewski}, \&
  {Cook}}]{Cote1999}
{C{\^o}t{\'e}}, P., {Mateo}, M., {Olszewski}, E.~W., \& {Cook}, K.~H. 1999,
  \apj, 526, 147

\bibitem[{{de Vaucouleurs}(1959)}]{deVaucouleurs1959}
{de Vaucouleurs}, G. 1959, Handbuch der Physik, 53, 275

\bibitem[{{Duc} {et~al.}(2015){Duc}, {Cuillandre}, {Karabal}, {Cappellari},
  {Alatalo}, {Blitz}, {Bournaud}, {Bureau}, {Crocker}, {Davies}, {Davis}, {de
  Zeeuw}, {Emsellem}, {Khochfar}, {Krajnovi{\'c}}, {Kuntschner}, {McDermid},
  {Michel-Dansac}, {Morganti}, {Naab}, {Oosterloo}, {Paudel}, {Sarzi}, {Scott},
  {Serra}, {Weijmans}, \& {Young}}]{Duc2015}
{Duc}, P.-A., {Cuillandre}, J.-C., {Karabal}, E., {et~al.} 2015, \mnras, 446,
  120

\bibitem[{{Forbes} {et~al.}(2012){Forbes}, {Cortesi}, {Pota}, {Foster},
  {Romanowsky}, {Merrifield}, {Brodie}, {Strader}, {Coccato}, \&
  {Napolitano}}]{Forbes2012}
{Forbes}, D.~A., {Cortesi}, A., {Pota}, V., {et~al.} 2012, \mnras, 426, 975

\bibitem[{{Forbes} {et~al.}(2011){Forbes}, {Spitler}, {Graham}, {Foster},
  {Hau}, \& {Benson}}]{Forbes2011LEDA}
{Forbes}, D.~A., {Spitler}, L.~R., {Graham}, A.~W., {et~al.} 2011, \mnras, 413,
  2665

\bibitem[{{Fried} \& {Illingworth}(1994)}]{Fried1994}
{Fried}, J.~W. \& {Illingworth}, G.~D. 1994, \aj, 107, 992

\bibitem[{{Gil de Paz} {et~al.}(2007){Gil de Paz}, {Boissier}, {Madore},
  {Seibert}, {Joe}, {Boselli}, {Wyder}, {Thilker}, {Bianchi}, {Rey}, {Rich},
  {Barlow}, {Conrow}, {Forster}, {Friedman}, {Martin}, {Morrissey}, {Neff},
  {Schiminovich}, {Small}, {Donas}, {Heckman}, {Lee}, {Milliard}, {Szalay}, \&
  {Yi}}]{GilDePaz2007}
{Gil de Paz}, A., {Boissier}, S., {Madore}, B.~F., {et~al.} 2007, \apjs, 173,
  185

\bibitem[{{Graham} {et~al.}(2012){Graham}, {Spitler}, {Forbes}, {Lisker},
  {Moore}, \& {Janz}}]{Graham2012}
{Graham}, A.~W., {Spitler}, L.~R., {Forbes}, D.~A., {et~al.} 2012, \apj, 750,
  121

\bibitem[{{Hao} {et~al.}(2006){Hao}, {Mao}, {Deng}, {Xia}, \& {Wu}}]{Hao2006}
{Hao}, C.~N., {Mao}, S., {Deng}, Z.~G., {Xia}, X.~Y., \& {Wu}, H. 2006, \mnras,
  370, 1339

\bibitem[{{Hubble}(1926)}]{Hubble1926}
{Hubble}, E.~P. 1926, \apj, 64, 321

\bibitem[{{Jedrzejewski}(1987)}]{Jedrzejewski1987}
{Jedrzejewski}, R.~I. 1987, \mnras, 226, 747

\bibitem[{{Kim}(1989)}]{Kim1989}
{Kim}, D.-W. 1989, \apj, 346, 653

\bibitem[{{Koch}(2009)}]{Koch2009Review}
{Koch}, A. 2009, Astronomische Nachrichten, 330, 675

\bibitem[{{Koch} {et~al.}(2017){Koch}, {Black}, {Rich}, {Longstaff}, {Collins},
  \& {Janz}}]{Koch2017_1661}
{Koch}, A., {Black}, C.~S., {Rich}, R.~M., {et~al.} 2017, Astronomische
  Nachrichten, 338, 503

\bibitem[{{Koch} {et~al.}(2012){Koch}, {Burkert}, {Rich}, {Collins}, {Black},
  {Hilker}, \& {Benson}}]{Koch2012HCC}
{Koch}, A., {Burkert}, A., {Rich}, R.~M., {et~al.} 2012, \apjl, 755, L13

\bibitem[{{Kroupa}(2001)}]{Kroupa2001}
{Kroupa}, P. 2001, \mnras, 322, 231

\bibitem[{{Lang} {et~al.}(2010){Lang}, {Hogg}, {Mierle}, {Blanton}, \&
  {Roweis}}]{Lang2010}
{Lang}, D., {Hogg}, D.~W., {Mierle}, K., {Blanton}, M., \& {Roweis}, S. 2010,
  \aj, 139, 1782

\bibitem[{{McConnachie}(2012)}]{McConnachie2012}
{McConnachie}, A.~W. 2012, \aj, 144, 4

\bibitem[{{McWilliam} \& {Zoccali}(2010)}]{McWilliam2010}
{McWilliam}, A. \& {Zoccali}, M. 2010, \apj, 724, 1491

\bibitem[{{Morrissey} {et~al.}(2007){Morrissey}, {Conrow}, {Barlow}, {Small},
  {Seibert}, {Wyder}, {Budav{\'a}ri}, {Arnouts}, {Friedman}, {Forster},
  {Martin}, {Neff}, {Schiminovich}, {Bianchi}, {Donas}, {Heckman}, {Lee},
  {Madore}, {Milliard}, {Rich}, {Szalay}, {Welsh}, \& {Yi}}]{Morrissey2007}
{Morrissey}, P., {Conrow}, T., {Barlow}, T.~A., {et~al.} 2007, \apjs, 173, 682

\bibitem[{{Mu{\~n}oz} {et~al.}(2015){Mu{\~n}oz}, {Eigenthaler}, {Puzia},
  {Taylor}, {Ordenes-Brice{\~n}o}, {Alamo-Mart{\'\i}nez}, {Ribbeck},
  {{\'A}ngel}, {Capaccioli}, {C{\^o}t{\'e}}, {Ferrarese}, {Galaz}, {Hempel},
  {Hilker}, {Jord{\'a}n}, {Lan{\c{c}}on}, {Mieske}, {Paolillo}, {Richtler},
  {S{\'a}nchez-Janssen}, \& {Zhang}}]{Munoz2015}
{Mu{\~n}oz}, R.~P., {Eigenthaler}, P., {Puzia}, T.~H., {et~al.} 2015, \apjl,
  813, L15

\bibitem[{{M{\"u}ller} {et~al.}(2015){M{\"u}ller}, {Jerjen}, \&
  {Binggeli}}]{Mueller2015}
{M{\"u}ller}, O., {Jerjen}, H., \& {Binggeli}, B. 2015, \aap, 583, A79

\bibitem[{{Park} {et~al.}(2017){Park}, {Moon}, {Zaritsky}, {Pak}, {Lee}, {Kim},
  {Kim}, \& {Cha}}]{Park2017}
{Park}, H.~S., {Moon}, D.-S., {Zaritsky}, D., {et~al.} 2017, \apj, 848, 19

\bibitem[{{Pasquali} {et~al.}(2007){Pasquali}, {van den Bosch}, \&
  {Rix}}]{Pasquali2007}
{Pasquali}, A., {van den Bosch}, F.~C., \& {Rix}, H.~W. 2007, \apj, 664, 738

\bibitem[{{Pastorello} {et~al.}(2014){Pastorello}, {Forbes}, {Foster},
  {Brodie}, {Usher}, {Romanowsky}, {Strader}, \& {Arnold}}]{Pastorello2014}
{Pastorello}, N., {Forbes}, D.~A., {Foster}, C., {et~al.} 2014, \mnras, 442,
  1003

\bibitem[{{Peng} {et~al.}(2002){Peng}, {Ho}, {Impey}, \& {Rix}}]{Peng2002}
{Peng}, C.~Y., {Ho}, L.~C., {Impey}, C.~D., \& {Rix}, H.-W. 2002, \aj, 124, 266

\bibitem[{{Pulsoni} {et~al.}(2018){Pulsoni}, {Gerhard}, {Arnaboldi}, {Coccato},
  {Longobardi}, {Napolitano}, {Moylan}, {Narayan}, {Gupta}, {Burkert},
  {Capaccioli}, {Chies-Santos}, {Cortesi}, {Freeman}, {Kuijken}, {Merrifield},
  {Romanowsky}, \& {Tortora}}]{Pulsoni2018}
{Pulsoni}, C., {Gerhard}, O., {Arnaboldi}, M., {et~al.} 2018, \aap, 618, A94

\bibitem[{{Rembold} {et~al.}(2005){Rembold}, {Pastoriza}, \&
  {Bruzual}}]{Rembold2005}
{Rembold}, S.~B., {Pastoriza}, M.~G., \& {Bruzual}, G. 2005, \aap, 436, 57

\bibitem[{{Rich} {et~al.}(2012){Rich}, {Collins}, {Black}, {Longstaff}, {Koch},
  {Benson}, \& {Reitzel}}]{Rich2012}
{Rich}, R.~M., {Collins}, M.~L.~M., {Black}, C.~M., {et~al.} 2012, \nat, 482,
  192

\bibitem[{{Rich} {et~al.}(2019){Rich}, {Mosenkov}, {Lee-Saunders}, {Koch},
  {Kormendy}, {Kennefick}, {Brosch}, {Sales}, {Bullock}, {Burkert}, {Collins},
  {Cooper}, {Fusco}, {Reitzel}, {Thilker}, {Milewski}, {Elias}, {Saade}, \& {De
  Groot}}]{Rich2019}
{Rich}, R.~M., {Mosenkov}, A., {Lee-Saunders}, H., {et~al.} 2019, \mnras, 490,
  1539

\bibitem[{{Schlafly} \& {Finkbeiner}(2011)}]{Schlafly2011}
{Schlafly}, E.~F. \& {Finkbeiner}, D.~P. 2011, \apj, 737, 103

\bibitem[{{Tal} {et~al.}(2009){Tal}, {van Dokkum}, {Nelan}, \&
  {Bezanson}}]{Tal2009}
{Tal}, T., {van Dokkum}, P.~G., {Nelan}, J., \& {Bezanson}, R. 2009, \aj, 138,
  1417

\bibitem[{{Torrealba} {et~al.}(2019){Torrealba}, {Belokurov}, {Koposov}, {Li},
  {Walker}, {Sanders}, {Geringer-Sameth}, {Zucker}, {Kuehn}, {Evans}, \&
  {Dehnen}}]{Torrealba2019}
{Torrealba}, G., {Belokurov}, V., {Koposov}, S.~E., {et~al.} 2019, \mnras, 488,
  2743

\bibitem[{{Vazdekis} {et~al.}(2016){Vazdekis}, {Koleva}, {Ricciardelli},
  {R{\"o}ck}, \& {Falc{\'o}n-Barroso}}]{Vazdekis2016}
{Vazdekis}, A., {Koleva}, M., {Ricciardelli}, E., {R{\"o}ck}, B., \&
  {Falc{\'o}n-Barroso}, J. 2016, \mnras, 463, 3409

\bibitem[{{Zanatta} {et~al.}(2018){Zanatta}, {Cortesi}, {Chies-Santos},
  {Forbes}, {Romanowsky}, {Alabi}, {Coccato}, {Mendes de Oliveira}, {Brodie},
  \& {Merrifield}}]{Zanatta2018}
{Zanatta}, E. J.~B., {Cortesi}, A., {Chies-Santos}, A.~L., {et~al.} 2018,
  \mnras, 479, 5124

\end{thebibliography}
\end{document}